\documentclass[useAMS,usenatbib]{mn2e}

\usepackage{graphicx,amssymb}
\usepackage[normalem]{ulem}
\usepackage[usenames,dvipsnames]{xcolor}
\usepackage{soul}

\title[Sesquinary catastrophe around WDs]
{Bounding destruction timescales of minor planets orbiting white dwarfs with the sesquinary catastrophe}

\author[]{Dimitri Veras$^{1,2,3}$\thanks{E-mail: dimitri.veras@aya.yale.edu},
Matija \'{C}uk$^{4}$,
\\
$^{1}$Centre for Exoplanets and Habitability, University of Warwick, Coventry CV4 7AL, UK
\\
$^{2}$Centre for Space Domain Awareness, University of Warwick, Coventry CV4 7AL, UK
\\
$^{3}$Department of Physics, University of Warwick, Coventry CV4 7AL, UK
\\
$^{4}$SETI Institute, 339 N Bernardo Avenue, Mountain View, CA 94043, USA
}

\begin{document}
\label{firstpage}
\pagerange{\pageref{firstpage}--\pageref{lastpage}}
\maketitle

\begin{abstract}
Dynamical activity attributed to the destruction of minor planets orbiting white dwarfs has now been photometrically monitored in individual systems for up to one decade, long enough to measure significant cessation and re-emergence of transit features. Further, periodicities which hint at the presence of debris orbiting exterior to the white dwarf Roche radius, along with widely varying estimates for debris disc lifetimes (up to Myrs), complicate theories for the formation and dynamical evolution of these systems. Here, we illustrate that minor planets orbiting white dwarfs with periods of $\approx$~5-25 hours and longer while completely or partially avoiding tidal disruption satisfy the conditions for the occurrence of the sesquinary catastrophe, a phenomenon that occurs in the solar system when impacts from returning ejecta from a moon are fast enough to be erosional to the point of destruction. We hence find that the region corresponding to $\approx$~1-4 white dwarf rubble-pile Roche radii represents a danger zone where the collisional timescale for the sesquinary catastrophe to occur is $\sim 10^2-10^5$~yr, suggesting that debris discs around white dwarfs are in a state of semi-continuous replenishment.
\end{abstract}


\section{Introduction}

The region interior to $0.01$~au of a white dwarf planetary system is a violent, extreme environment. Featuring planetary debris which all originated from much further away, orbital velocities in excess of 300 km s$^{-1}$, discs and rings with dust and gas, and a lot of space between the Earth-sized white dwarf and its Solar-sized Roche sphere boundary, this region is a laboratory for a plethora of physical processes.

The possibility of the disruption of minor planets close to white dwarfs has been theorised for decades \citep{graetal1990,debsig2002,jura2003}, but not realised in direct photometric observations until 2015 \citep{vanetal2015}. Now, we assume that this process is effectively ubiquitous, helping to generate the metal enrichment seen in over 1700 white dwarfs \citep{wiletal2024}, representing about a quarter of all currently known exoplanetary systems. Dedicated surveys indicate that 15-50 per cent of all Milky Way white dwarfs contain planetary remnants \citep{zucetal2010,koeetal2014,obretal2023}, and on the order of 100 discs of planetary dust or gas have been detected through spectroscopy and spectral energy distributions \citep[e.g.][]{zucbec1987,ganetal2006,faretal2009,manetal2020,xuetal2020,swaetal2024,faretal2025,rogetal2025}.

The photometric observational signatures of minor planet disruption around white dwarfs are represented as transit curves of the resultant dusty and gaseous effluences. Unlike for exoplanet transits, which are symmetrical and have typical depths of under a few per cent, the disruption of minor planets around white dwarfs produces highly intricate repeating and non-repeating asymmetric structures in transit curves, with depths sometimes in excess of 50 per cent \citep[e.g.][]{ganetal2016,garetal2017,kjuetal2017,izqetal2018,rapetal2018}. Such large depths indicate that the structures blocking the star's radiation are dust and gas clouds rather than the minor planets themselves.

Currently on the order of 10 white dwarfs have observed transiting debris \citep{vanetal2015,vanbosetal2020,vanbosetal2021,faretal2022,bhaetal2025,heretal2025}, from which at least five distinct orbital periodicities have been measured (4.49~hr, 9.94~hr, 14.8~hr, 25.0~hr, 107~d). The smallest of these values corresponds to debris which resides closest to the rubble-pile Roche limit. This initially prominent 4.49~hr feature from the now well-known WD~1145+017 planetary system was the first periodicity to be measured \citep{vanetal2015} and has now been shown to have effectively disappeared within one decade \citep{aunetal2024}.

A decadal timescale is orders of magnitude shorter than the white dwarf debris disc lifetimes that have been observationally inferred from \cite{giretal2012} and \cite{cunetal2025}, and is at the lower end of the timescales analytically computed from \cite{verhen2020}. Other efforts to compute stability timescales in the immediate vicinity of white dwarfs have involved $N$-body simulations with equal-mass co-orbital objects close to the Roche radius \citep{veretal2016} and the circular restricted three-body problem with extreme mass ratios appropriate for asteroid-white dwarf systems \citep{antver2024}. Complicating discussions of debris lifetimes are the different ways minor planets can break up, especially just outside of the white dwarf's Roche sphere \citep{makver2019,veretal2020,sheser2023}, plus the myriad of disc-based physics affecting evolutionary processes (see \citealt*{veras2021}, \citealt*{malamud2024} and \citealt*{veretal2024} for reviews).

One prominent potential destruction process is collisions, and investigations which were dedicated to collisional cascade modelling include \cite{kenbro2017a}, \cite{kenbro2017b} and \cite{lietal2021}. Cratering on asteroidal or moon-sized objects has not been as much of a focus; \cite{verkur2020} computed the ejecta mass from cratering events, where the ejecta represents a way to replenish stagnant debris reservoirs around white dwarfs. However, they did not track the stellocentric trajectories of the debris nor consider the possibility of returning ejecta.

In the solar system, impact ejecta which escapes a moon on a roughly similar orbit may later impact the same moon, producing sesquinary craters. Bolstered by evidence from crater morphologies, this phenomenon is thought to be common \cite[e.g.][]{alvetal2005,alvetal2008,zahetal2008,bieetal2012,nayasp2016,nayetal2016,alvetal2017}. The outcome of sesquinary impacts is typically reaccretion onto the moon. However, \cite{cuketal2023} showed that under certain conditions, the velocity of the returning ejecta is high enough to be erosional, potentially leading to total breakup and possible reaccretion in a ``sesquinary catastrophe".

Here we show that minor planets orbiting white dwarfs within at least several Roche radii satisfy the conditions for the sesquinary catastrophe to occur. Further, the ejecta do not need to be generated exclusively or even partly from initial impacts. We are not claiming that this process can explain the complex diminution of transit depths over time of any one system, nor that we know the detailed state of the minor planet or its remnants during or after the catastrophe. However, the existence of the phenomena in white dwarf planetary systems allows us to usefully bound the destruction timescale of the minor planet as a function of both its orbital period (or distance from the Roche radius) and its orbital parameters by using the collisional timescale as a proxy.

The paper is structured as follows. In Section 2, we describe the parameter space that is the focus of our study (minor planet orbital periods of 5-25 hours), and make the link with white dwarf Roche radii. Then, in Section 3, we outline the conditions required for the sesquinary catastrophe to occur. These conditions include a minimum level of eccentricity or inclination excitation, as well as the existence of differential apsidal or nodal precession. In Section 4, we then characterise the timescale of the sesquinary catastrophe. We finally discuss our results in Section 5 before concluding in Section 6.

\section{Relevant parameter space}

The types of objects which are commonly assumed to disrupt around a white dwarf are strengthless rubble piles, an apt description for many solar system asteroids. A white dwarf's rubble-pile Roche radius, $r_{\rm Roche}$, can be given by

\begin{equation}
r_{\rm Roche} = k \left(\frac{M_{\rm WD}}{\rho_{\rm mp}}\right)^{1/3}
\end{equation}

\noindent{}where $M_{\rm WD}$ is the mass of the white dwarf, $\rho_{\rm mp}$ is the bulk density of the minor planet that is being disrupted, $k=0.78$ if the minor planet is not spinning, and $k=0.89$ if the minor planet is spinning synchronously with the white dwarf \citep{veretal2017}. 

In order to compare $r_{\rm Roche}$ with the range of observed periodicities in the immediate vicinity of the white dwarf (excepting the 107 day periodicity; see Section 5.1), we are able to use Kepler's third law in order to obtain at least the semi-major axis because the masses of the host white dwarf hosts are known. We list these masses, along with the associated periodicities, here:

\begin{itemize}

\item {\bf WD~1145+017} has an observed periodicity of $4.49$~hr and a host star mass of $M_{\rm WD}=0.66M_{\odot}$ \citep{vanetal2015,foretal2020}.

\item {\bf ZTF~J0328-1219} has an observed periodicity of $9.94$~hr and a host star mass of $M_{\rm WD}=0.73M_{\odot}$ \citep{vanbosetal2021}.

\item {\bf SBSS~1232+563} has an observed periodicity of $14.8$~hr and a host star mass of $M_{\rm WD}=0.77M_{\odot}$ \citep{couetal2019,heretal2025}.

\item {\bf WD~1054-226} has an observed periodicity of $25.0$~hr and a host star mass of $M_{\rm WD}=0.62M_{\odot}$ \citep{faretal2022}.


\end{itemize}

\begin{figure}
\includegraphics[width=8cm]{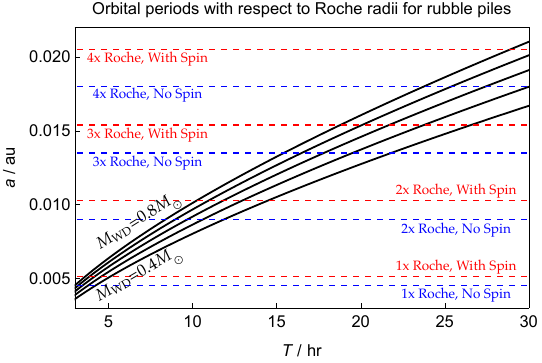}
\caption{
Semi-major axis $a$ versus orbital period $T$ for a minor planet around a white dwarf with a mass given by, from the top to bottom solid black curve, $M_{\rm WD} = 0.8M_{\odot}, 0.7M_{\odot}, 0.6M_{\odot}, 0.5M_{\odot}$, and $0.4M_{\odot}$. The bottom two horizontal dashed lines correspond to the rubble-pile Roche radii $r_{\rm Roche}$ of a $0.65M_{\odot}$ white dwarf for a minor planet with density $\rho_{\rm mp}=2$~g~cm$^{-3}$ which is not spinning (blue) and is spinning synchronously with the white dwarf (red). The other horizontal lines are multiples of $r_{\rm Roche}$. The $x$-axis encompasses the range of the majority of securely measured periodicities in photometric transits of white dwarf debris systems, and this plot illustrates approximately how far away a body can be from $r_{\rm Roche}$ while exhibiting the observed period.
}
\label{rocheplot}
\end{figure}

Although we do not necessarily assume that the observed periodicities exactly equal the orbital period, $T$, of the minor planet -- or whatever is left of it -- we do assume that, based on the above values, $T\approx 5-25$~hr is a reasonable range to explore. This choice is also supported with theoretical considerations of the tidally-induced orbit shrinkage of planetesimals to this region \citep{ocolai2020,lietal2025a,lietal2025b}. Hence, in Fig. \ref{rocheplot}, we plot a minor planet's semi-major axis $a$ versus $T$ for a range of white dwarf masses which encompass those in the above list, and overplot $r_{\rm Roche}$ for the $M_{\rm WD}=0.65M_{\odot}$ and $\rho=2$~g~cm$^{-3}$ case\footnote{The white dwarf masses in the bulleted list are on the slightly high end relative to the known Milky Way distribution. For single white dwarfs, either $0.60M_{\odot}$ or $0.65M_{\odot}$ represents a reasonable fiducial value \citep[e.g.][]{mccetal2020}.}.

The plot illustrates that periodicities above about 5 hours correspond to semi-major axes beyond the rubble-pile Roche radius, and that periodicities of tens of hours correspond to many times this disruption limit. Understanding what is happening in this region beyond the rubble-pile Roche radius is a key motivation for this work.

\section{Sesquinary catastrophe conditions}

Having now honed in on the most relevant parameter space, we can describe the conditions for the sesquinary catastrophe to occur. It does so when the relative velocity between the minor planet and its returning ejecta is sufficiently high to be erosional. For low relative velocities, the ejecta may just gently reaccrete, and hence not contribute to the minor planet's eventual destruction.

In order to attain a high-enough relative velocity, the orbits of the minor planet and its ejecta must be (i) eccentric and apsidally misaligned, or/and (ii) inclined and nodally misaligned. Before proceeding, we need to establish the plausibility of each scenario in white dwarf planetary systems.

\subsection{Orbit misalignment}

Apsidal and nodal precession occur in every planetary system that features non-circular and inclined orbits, just on different timescales depending on the system architecture. We need to estimate these timescales. To do so, and in anticipation of future discoveries, we include a variety of potential effects, and in the following order: stellar oblateness, magnetic forces, additional bodies (or structures, such as an existing disc) in the system, and general relativity.

The nodal precession rate, which is technically a regression rate, can be approximated as \citep{broken2022,freetal2025}

\[
\dot{\Omega} \sim -\sqrt{\frac{GM_{\rm WD}}{a^3}}
                            \bigg[\frac{3}{2}J_2\left(\frac{R_{\rm WD}}{a}\right)^2 
                            +\frac{9B_{\rm WD}^2R_{\rm WD}^6}{G M_{\rm WD} \rho_{\rm mp} a^5}
\]

\begin{equation}
\ \ \ \ \ \ \ \ \
+ \frac{1}{4}\sum_{j} \frac{M_{j}}{M_{\rm WD}} \alpha_{j}^2 b_{3/2}^{(1)}\left( \alpha_{j} \right)
\bigg],
\label{nodeEQ}
\end{equation}

\noindent{}where $R_{\rm WD}$ is the radius of the white dwarf, $B_{\rm WD}$ is the magnetic field strength of the white dwarf, $J_2$ is a dimensionless parameter which indicates the white dwarf's oblateness to quadrupole order, $M_{j}$ is another object in the system with label $j$, $\alpha_j = a/a_j$, and $b_{3/2}^{(1)}(\alpha_j)$ is a (dimensionless) Laplace coefficient evaluated at $\alpha_j$. 

The apsidal precession rate is

\[
\dot{\omega} \sim \sqrt{\frac{GM_{\rm WD}}{a^3}}
                            \bigg[\frac{3}{2}J_2\left(\frac{R_{\rm WD}}{a}\right)^2 
                            +\frac{45B_{\rm WD}^2R_{\rm WD}^6}{2G M_{\rm WD} \rho_{\rm mp} a^5}
\]

\begin{equation}
\ \ \ \ \ \ \ \ \
+ \frac{1}{4}\sum_{j} \frac{M_{j}}{M_{\rm WD}} \alpha_{j}^2 b_{3/2}^{(1)}\left( \alpha_{j} \right)
+ \frac{3GM_{\rm WD}}{a c^2 \left(1-e^2\right)}
\bigg],
\label{apsEQ}
\end{equation}

\noindent{}where $c$ is the speed of light. A comparison of equations (\ref{nodeEQ}) and (\ref{apsEQ}) illustrates that general relativity has no effect on nodal precession, and the magnitudes of the other three effects are nearly equivalent, at least in this approximation. Further, oblateness and general relativity are present in every white dwarf system, whereas magnetism and additional objects may not be.

\subsubsection{Oblateness}

Let us now consider each effect in turn, starting with oblateness. In the solar system, moons of the giant planets, and in particular Saturn, experience quick apsidal and nodal precession because the value of $J_2$ is so large; for Saturn, $J_2\approx 0.016$ \citep{iesetal2019}. As a result, oblateness is often the dominant driver of misaligned orbits, and what helps to allow for the sesquinary catastrophe to occur.

Hence, in order to determine if a similar situation holds for white dwarfs, we need to compute a representative $J_2$ value for a white dwarf. To do so, we can relate $J_2$ to a white dwarf's structure and spin through \citep{sterne1939,waretal1976,spaetal2018}

\begin{equation}
J_2 = \frac{k_2 R_{\rm WD}^3}{3GM_{\rm WD}} S_{\rm WD}^2,
\label{J2eq}
\end{equation}

\noindent{}where $k_2$ is the white dwarf's Love number and $S_{\rm WD}$ is the spin rate of the white dwarf. From Fig. 2 of \cite{batada2013}, $k_2 = 0.02$ or $0.28$ for a fully radiative or fully convective star, respectively; these values can be considered bounding cases. Further, recent data for white dwarf rotation reveals a potentially four-orders-of-magnitude range in spin rate, from one revolution per 0.1 hours to one revolution every 100 hours \citep{olietal2024}.

In addition to $k_2$ and $S_{\rm WD}$, we also require representative values of $M_{\rm WD}$ and $R_{\rm WD}$. Because we are computing ranges, we consider $0.4M_{\odot} \le M_{\rm WD} \le 0.8M_{\odot}$ and compute the corresponding radii according to \citep{nauenberg1972}

\begin{equation}
\frac{R_{\rm WD}}{R_{\odot}} \approx 0.0127 \left(\frac{M_{\rm WD}}{M_{\odot}} \right)^{-1/3}
\sqrt{1-0.607 \left(\frac{M_{\rm WD}}{M_{\odot}} \right)^{4/3}},
\end{equation}

\noindent{}yielding $0.010R_{\odot} \lesssim R_{\rm WD} \lesssim 0.016R_{\odot}$, where the higher radius bound corresponds to the lower mass bound.

Inserting all of these values into equation (\ref{J2eq}) yields a substantial range, $J_2 \sim 10^{-13}-10^{-5}$, rather than any single representative order-of-magnitude. Regardless, even when assuming the highest possible $J_2$ value, then the timescale for either nodal or apsidal precession ($=2\pi/\dot{\Omega}$ and $2\pi/\dot{\omega}$) is greater than $\sim 10^6$~yr, which is too long to be relevant in white dwarf debris disc systems.

\subsubsection{Magnetic fields}

The next effect to consider is magnetic fields. Only tens of per cent of white dwarfs actually feature detectable magnetic fields with $B_{\rm WD} \gtrsim 0.1~{\rm T}$ ($10^3$~G) \citep[e.g.][]{feretal2015}, and the highest observed magnetic fields are on the order of $B_{\rm WD} \sim 10^5~{\rm T}$ ($10^9$~G) \citep[e.g.][]{brietal2018}. Even in the highest $B_{\rm WD}$ case, the timescale for magnetically-induced nodal or apsidal precession at $\approx 0.01$~au is on the order of a Hubble time, and hence is not relevant here\footnote{The steep dependence of the magnetic precession timescale on $a$ indicates that this type of precession would become important for minor planets with high internal strength on eccentric orbits at distances well within the rubble-pile Roche radius, e.g. in the SDSS~J1228+1040 white dwarf planetary system, if only the host star was more magnetic \citep{manetal2019,treetal2021}. \cite{broken2022} showed that the precession rate for minor planets orbiting magnetic white dwarfs is many degrees per year at $a\approx0.2R_{\odot}\approx 10^{-3}$~au.}.

\subsubsection{Other objects in the system}

White dwarfs cool quickly and become dim, challenging efforts to detect planetesimals, planets and other orbiting objects. As a result, there might be massive bodies lurking just outside of our detection windows, even in systems that contain observable transiting debris. Further, of the handful of known white dwarf planets, two of these are giant planets that reside entirely within 0.1~au of their respective white dwarf hosts \citep{ganetal2019,WD1856}. Those systems do not host transiting debris (yet), but it is not unreasonable to assume a future discovery of a white dwarf planetary system featuring both a planet and transiting debris \citep[e.g.][]{debetal2025}.

If there is a giant planet (or something even more massive) lurking within 0.1~au of a white dwarf planetary system, then equations (\ref{nodeEQ})-(\ref{apsEQ}) show that the planet will generate {\it both} nodal and apsidal precession on relevant timescales. These timescales range from $\sim 10^2-10^3$~yr for a planet/star mass ratio of $10^{-3}$, $a=0.01$~au and $\alpha=1/9-1/4$. Further, because the precession rates scale linearly with planet mass, a mini-Neptune or Super-Earth would generate misalignment timescales which are only a few orders of magnitude longer. 

Other minor planets or a debris disc currently in the system are unlikely to be massive enough to create significant differential precession. By comparison with the Saturnian system, while the numerator in the mass ratio in the summation in equations (\ref{nodeEQ})-(\ref{apsEQ}) might be similar, the denominator in white dwarf systems is 3-4 orders of magnitude higher.

\subsubsection{General relativity}

\begin{figure}
\includegraphics[width=8cm]{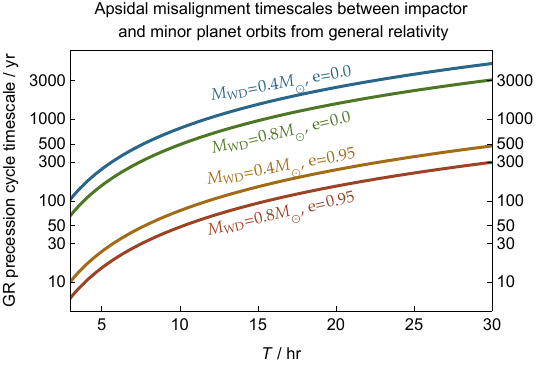}
\caption{
The apsidal misalignment timescale as a function of the minor planet's orbital period $T$ and its orbital eccentricity $e$. In most cases, these timescales approximate the minimum timescale over which the sesquinary catastrophe can occur.
}
\label{GRplot}
\end{figure}

General relativity generates apsidal precession in every white dwarf planetary system and nodal precession in none of them. Further, the magnitude of the general relativity precession term in equation (\ref{apsEQ}) is high enough to be the most relevant precession term in almost all cases. In order to quantify this notion, we plot the apsidal precession timescale due to general relativity in Fig. \ref{GRplot}.

The plot reveals precession timescales that are comparable to those which would be generated by the presence of a nearby giant planet. Given the current lack of such planets, we conclude that general relativity provides the minimum timescale for the sesquinary catastrophe to occur through apsidal misalignment of eccentric orbits in nearly all cases. Alternatively, the presence of a nearby giant planet would set the minimum timescale for the sesquinary catastrophe to occur either through nodal precession on inclined orbits, or/and through apsidal precession on eccentric orbits, especially for more distant ($T \gtrsim 30$~hr) minor planets where the influence of general relativity is weak.

\subsection{Orbital excitation}

Our analysis from the last subsection indicates that with regards to the sesquinary catastrophe, the minor planet's eccentricity is always important, whereas its inclination may be important in a few select cases (with other massive objects in the system, hidden or not). Hence, we proceed by considering the full three-dimensional case for orbital excitation, even though for the currently observed set of transiting systems, we can reasonably just set $i=0^{\circ}$ (where $i$ is the minor planet's inclination relative to the stellar equator) and only focus on the resulting equations and plot curves.

In order for the sesquinary catastrophe to occur, the relative velocity between the minor planet and a returning impactor must be sufficiently high. \cite{cuketal2023} approximate this critical velocity through a dimensionless parameter $q_v$ in terms of primarily minor planet properties as

\begin{equation}
q_{v} = \sqrt{e^2 + \sin^2{i}} \left(\frac{v_{\rm orb}}{v_{\rm esc}} \right).
\label{qveq}
\end{equation}

The escape velocity from the minor planet is

\begin{equation}
v_{\rm esc} = \sqrt{\frac{2GM_{\rm mp}}{R_{\rm mp}}},
\end{equation}

\noindent{}where $M_{\rm mp}$ and $R_{\rm mp}$ are the mass and radius of the minor planet; the minor planet is assumed to be spherical, which is a good enough approximation for this broad investigation. 

The orbital velocity $v_{\rm orb}$ of the minor planet is 

\begin{equation}
v_{\rm orb} \approx \sqrt{G M_{\rm WD} \left(\frac{2}{r} - \frac{1}{a}\right)},
\end{equation}

\noindent{}where $r$ is the distance between the white dwarf and minor planet

\begin{equation}
r = \frac{a\left(1 - e^2\right)}{1 + e \cos{f}}
\end{equation}

\noindent{}with $f$ representing the true anomaly of the minor planet. Because equation (\ref{qveq}) is meant to be time-independent, in the following calculations, we set $f=0$ so that the value of $r$ represents the pericentre distance.

Note that $q_v = 0$ when $e=0$ and $i=0^{\circ}$, illustrating that perfectly co-orbital and coplanar impact ejecta will not be erosional because there won't be any differential orbital precession. Further, for $q_v>0$, the ejecta become erosional only when $q_v \gtrsim 10$, as deduced by \cite{cuketal2023}, based on the collisional properties from \cite{stelei2012}.

\begin{figure}
\includegraphics[width=8cm]{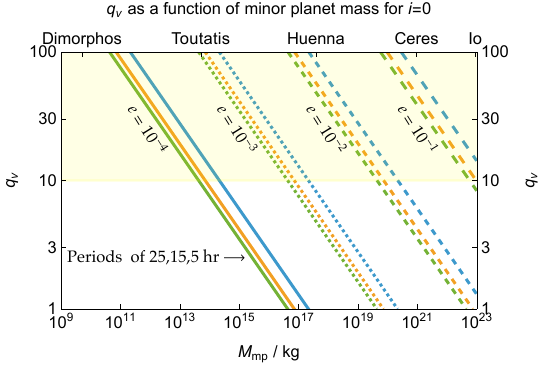}
\centerline{}
\includegraphics[width=8cm]{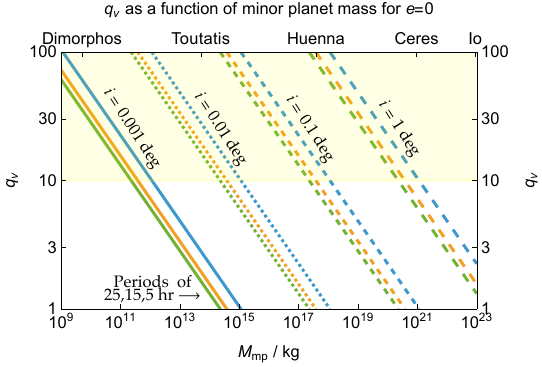}
\caption{
Values of the minor planet mass for which the sesquinary catastrophe occurs around white dwarfs ($q_v \gtrsim 10$). The top panel models the case where differential apsidal precession creates the catastrophe (true for nearly all white dwarf planetary systems), and the bottom panel models the case where differential nodal precession creates or contributes to the catastrophe (only true when a giant planet is close to the debris). Each plot shows curves for orbital periods of $T=5,15,25$~hr, approximately encompassing the periodic transit signals seen in the WD~1145+017, ZTF~J0328-1219, SBSS~1232+563 and WD~1054-226 white dwarf planetary systems. In all cases, the white dwarf mass is taken to be $0.65M_{\odot}$, the density of the minor planet is taken to be $\rho_{\rm mp}=2$~g\,cm$^{-3}$, and the minor planet is assumed to be spherical; the solar system objects on the top $x$-axes indicate only mass. Overall, both plots show that the sesquinary catastrophe is easily achieved for minor planets orbiting white dwarfs, even when the orbit is just slightly dynamically excited.
}
\label{gvplots}
\end{figure}

We can now evaluate $q_v$ for white dwarf systems. As shown in Section 2, we have precise values for $M_{\rm WD}$ and can use the measured periodicities as proxies for $T$. In turn, we can obtain values for $a$, but not for $e$ or $i$. These last two orbital parameters are not as well constrained. Nevertheless, theoretical work has provided some insight, particularly for WD~1145+017, a system for which over 20 dedicated papers have been published. 

Although the destruction of the minor planet in this system is primarily caused by tidal fragmentation rather than collisional processes (because $T=4.49$~hr is just at the rubble pile Roche radius), summarising the constraints that have been placed on $e$ may be useful. During the early epochs of its dynamical activity, the photometric transit curves for the debris orbiting WD~1145+017 featured multiple periodicities. Some of these were attributed to broken-off fragments of the minor planet \citep{rapetal2016}. \cite{guretal2017} then used $N$-body simulations to restrict the mass and eccentricity of the parent minor planet that would be perturbing these fragments, and found that $e\sim 10^{-3}-10^{-1}$ (with $M_{\rm mp} \approx 0.1M_{\rm Ceres}$). Numerical simulations of the actual breakup of the minor planet \citep{veretal2017} reinforced the constraint $e \lesssim 0.1$, although some features of the transit curves at singular moments in time could also be reproduced with exactly circular orbits \citep{duvetal2020}.

For the other three white dwarfs on the bulleted list in Section 2 (ZTF~J0328-1219, SBSS~1232+563, WD~1054-226), the value of $e$ can be as high as many tenths while avoiding the rubble-pile Roche limit. Further, the one other white dwarf for which we have measured periodicities, ZTF~J0139+5245, has a whopping transit period of 107 days \citep{vanbosetal2020}, meaning in this case $e$ could easily be as high as 0.95 and still avoid the Roche radius. 

Constraints on $i$ with respect to other objects in the system are arguably poorer given that hidden bodies could exist at any inclination outside of the narrow transit detectability window. Unlike dust and pebbles, which can be radiatively dragged into hot white dwarfs from distances of tens of au \citep{veretal2022}, minor planets require being gravitational scattered inwards from au-scale distances or larger. The inclination distribution of minor planets scattered into the close vicinity of white dwarfs was specifically investigated by \cite{veretal2021}, who found that the level of inclination isotropy increases with the mass of the assumed-to-be-single perturber. Further, \cite{musetal2018} found that with three massive perturbers, the inclination distribution of the scattered minor planets is broadly isotropic. An alternative approach to constrain $i$ is to use radiative transfer modelling to assess the geometry of the dust in the debris disc; \cite{baletal2022} did so for the G29-38 white dwarf planetary system (which does not contain transiting debris) and found that the disc is not flat.

These weak constraints on $e$ and $i$ lead us to choose a broad range of values that illustrate where the $q_v = 10$ threshold is crossed. We also need to specify parameters for the minor planet, and we select a generous range of $M_{\rm mp}$ from $10^9$~kg to $10^{23}$~kg, encompassing objects such as Dimorphos, Toutatis, Huenna (a typical Main Belt asteroid), Ceres, Pluto and Io; disc masses larger than Io's would be considered unusually massive (e.g. \citealt*{vanetal2018}). In all cases, we assume that the minor planet is spherical and has a bulk density of $\rho_{\rm mp} = 2$~g cm$^{-3}$.

Our results are plotted in Fig. \ref{gvplots}. They demonstrate, e.g., that a canonical minor planet mass equalling 10 per cent of that of Ceres needs an orbit which deviates from the circular case by an eccentricity of about just 0.01 or more in order for the sesquinary catastrophe to occur. For smaller masses, the orbital eccentricity excitation threshold is even lower. Unusually large minor planets comparable to Io could also experience the catastrophe if their eccentricity exceeds 0.1. Given that the vast majority of minor planets orbiting white dwarfs are likely much smaller than Io \citep{wyaetal2014}, the sesquinary catastrophe in white dwarf systems could be commonplace.

\section{Collisional timescale}

The criterion $q_v \gtrsim 10$ just demonstrates that the sesquinary catastrophe would occur if enough time passed, but does not specify a timescale. There must be enough time for the collisional cascade\footnote{We use this term, {\it collisional cascade}, in a different sense to the evolution presented in \cite{kenbro2017a}, \cite{kenbro2017b}, \cite{lietal2021} and \cite{broetal2022}. The sesquinary catastrophe is focussed on a single parent body and collisions between it and its ejecta, as opposed to mutual collisions between the ejecta or large fragments of the parent body. Further, for the sesquinary catastrophe to take place, the parent body needs to avoid tidal disruption, at least partly, as well as avoid a giant impact (i.e. destruction due to the impact of a similarly large body, presumably incoming from the exterior regions of the planetary system).} to actually propagate and cause the destruction. \cite{cuketal2023} approximate this timescale as a few times $\tau$, where

\[
\tau \approx T\left(\frac{v_{\rm orb}}{v_{\rm esc}}\right) \left(\frac{a}{R_{\rm mp}}\right)^2 
\]

\begin{equation}
\ \ \ \ \ \
\times \sqrt{\left[ \left(\frac{v_{\rm esc}}{v_{\rm orb}}\right)^2 + e^2 \right]
             \left[ \left(\frac{v_{\rm esc}}{v_{\rm orb}}\right)^2 + \sin^2{i} \right]}.
\end{equation}

\noindent{}Because we are only interested in $\tau$ when $q_v \gtrsim 10$, we would like to determine the functional dependence of $\tau$ on $q_v$. We obtain

\[
\tau = \left(\frac{2}{3\pi}\right) G T^3 \rho_{\rm mp} 
\]

\[
\
\times 
q_{v}
\left( \frac{1-e}{1+e} \right)
\sqrt{
\frac
{\left[ e^2\left(1 + q_{v}^2\right) + \sin^2{i} \right]
 \left[ e^2 + \left(1 + q_{v}^2\right)\sin^2{i} \right] }
{\left(e^2 + \sin^2{i} \right)^3}
}.
\]

\begin{equation}
\label{tauqv}
\end{equation}

\begin{figure}
\includegraphics[width=8cm]{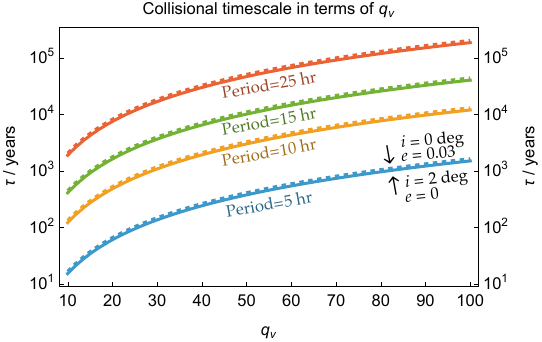}
\caption{
The collisional timescale $\tau$ as a function of the threshold for the sesquinary catastrophe to occur ($q_v \gtrsim 10$). The dependence is monotonic, and is a strong ($\propto T^3$) function of orbital period (equation \ref{tauqv}). The solid lines correspond to $i=2^{\circ}$ and $e=0$, whereas the dashed lines correspond to $i=0^{\circ}$ and $e=0.03$.
}
\label{tauplot1}
\end{figure}

\begin{figure}
\includegraphics[width=8cm]{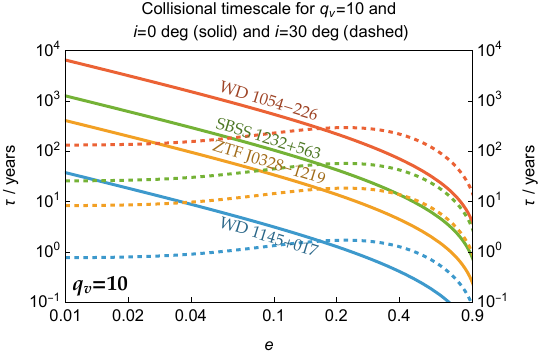}
\centerline{}
\includegraphics[width=8cm]{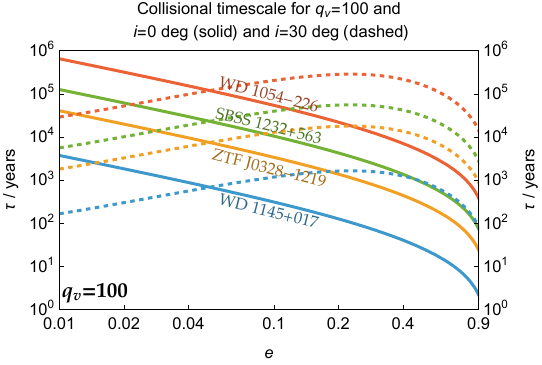}
\caption{
The collisional timescale $\tau$ for minor planets which are assumed to reside on orbits with the same main periodicities observed in the WD~1145+017, ZTF~J0328-1219, SBSS~1232+563 and WD~1054-226 white dwarf planetary systems. The upper and lower panels highlight the difference between the $q_v=10$ and $q_v=100$ cases. The solid lines correspond to $i=0^{\circ}$ and the dashed lines correspond to $i=30^{\circ}$. These curves are independent of white dwarf mass and minor planet mass and radius, but do assume $\rho_{\rm mp}=2$~g~cm$^{-3}$. The overall low values of $\tau$, at least compared to Myr timescales, lend more credence to the idea that debris discs around white dwarfs are predominantly in a state of semi-continuous replenishment. 
}
\label{tauplot23}
\end{figure}

For the apsidal precession case only ($i=0^{\circ}$), and for both small and large values of $e$, equation (\ref{tauqv}) reduces to

\begin{equation}
\tau_{\left(i=0^{\circ},e\approx 0\right)} = 
\left(\frac{2}{3\pi}\right) G T^3 \rho_{\rm mp}
\left(
\frac{1 - 2e + 2e^2}{e}
\right)
\sqrt{1 + q_{v}^2},
\label{Lowe}
\end{equation}

\begin{equation}
\tau_{\left(i=0^{\circ},e\approx 1\right)} = 
\left(\frac{1}{3\pi}\right) G T^3 \rho_{\rm mp}
\left(1-e\right)\sqrt{1 + q_{v}^2}.
\label{Highe}
\end{equation}

\noindent{}These two cases physically correspond to a minor planet at the late and early stages of tidally circularising from an initially scattered orbit. We use equation (\ref{tauqv}) as a rough proxy for the destruction timescale (generated by the sesquinary catastrophe only) for a minor planet orbiting a white dwarf when $q_v \gtrsim 10$. 

In order to illustrate how $\tau$ scales with $q_v$, we plot the equation in Fig. \ref{tauplot1}. On the figure, we choose values of $i$ and $e$ such that the curves nearly coincide. The plot then illustrates that $\tau$ scales monotonically with $q_v$, a dependence that can also be gleaned directly from any of the equations (\ref{tauqv}), (\ref{Lowe}) or (\ref{Highe}). Further, for a given $q_v$, the range $T=5-25$~hr generates a range of over two orders of magnitude in $\tau$, given that $\tau \propto T^3$

For a more direct application to the white dwarf systems with measured periodicities of transiting debris, we can plot equation (\ref{tauqv}) for specific values of both $q_v$ and $T$, as a function of the (relatively unknown) values of $e$ and $i$. We do so in Fig. \ref{tauplot23}. Note that the curves in this figure are independent of both $M_{\rm WD}$ and $M_{\rm mp}$, from equation (\ref{tauqv}).

The figure illustrates how one can estimate a destruction timescale if one knows, or has constraints, on $e$ and $i$, and assumes that the minor planet being destroyed is associated with a specific periodicity. Consider, for example, ZTF~J0328-1219, which features a periodicity which easily allows for an $e\approx0.2$ orbit that avoids the rubble pile Roche radius of the white dwarf. If $q_v=100$, then for $i=0^{\circ}$, we obtain $\tau\approx 10^3$~yr, which is about twice the apsidal precession timescale from Fig. \ref{GRplot}. This value of $q_v$ would in turn require the minor planet mass to be in-between that of Ceres and Io.

The observed cessation timescale in WD~1145+017 ($< 10$~yr) is much smaller than plausible values of $\tau$, suggesting that $\tau$ provides an upper bound only on the destruction timescales when tidal fragmentation is avoided. In this respect, note that the upper plot in Fig. \ref{tauplot23} showcases the limiting case of $q_v=10$, where $\tau$ takes on the smallest possible values, corresponding to the largest possible $M_{\rm mp}$ values from Fig. \ref{gvplots}.

\section{Discussion}\label{discussion}

\subsection{Application to ZTF~J0139+5245}

The broad applications of this work are focussed on minor planets whose orbits are outside of the rubble-pile Roche radius by a factor of several (with $T\approx5-25$~hr). However, what about ZTF~J0139+5245, with a dominant periodicity of $T=107$~days? The value of $e$ in this system has been theorised to be in excess of 0.97 by assuming that the minor planet survived the giant branch phases of evolution (at au-scales) and has now reached the rubble-pile Roche sphere. The suggestion from both Figs. \ref{tauplot1} and \ref{tauplot23} is that for values of $e>0.9$, $\tau$ may be low for systems like ZTF~J0139+5245.

\begin{figure}
\includegraphics[width=8cm]{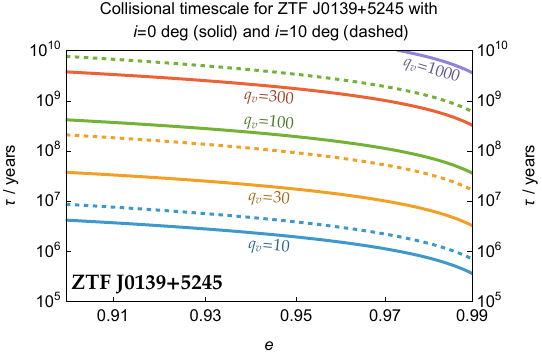}
\caption{
The collisional timescales for a minor planet on a highly eccentric orbit ($e>0.9$) breaking up due to the sesquinary catastrophe ($q_v \gtrsim 10$) in the ZTF~J0139+5245 system, where reaching the rubble pile Roche radius with $T=107$ days from au-scale distances would require $e \gg 0.9$. In almost all cases, the collisional timescale yields upper bounds which exceed even the highest estimates for debris discs lifetimes of white dwarfs.
}
\label{tauplot4}
\end{figure}

Hence, we explore the relevant timescales in this system. First, we compute the apsidal precession timescale due to general relativity, assuming $M_{\rm WD}=0.52M_{\odot}$ \citep{bedetal2020} and $e=0.95$. The resulting timescale is $\approx 0.6$~Myr. We then compute the collisional timescales for ZTF~J0139+5245 in Fig. \ref{tauplot4}. The plot illustrates that the values of $\tau$ are actually nearly all higher than $10^6$~yr, exceeding the disc lifetimes upper bounds given by \cite{giretal2012} and \cite{cunetal2025}. Hence, for long-period systems such as ZTF~J0139+5245, the sesquinary catastrophe may not provide as useful of a restriction when compared to disc lifetimes, but rather may be of more interest when comparing to the $\sim 0.01-10$~Gyr cooling age of a white dwarf.

\subsection{Origin of mass ejecta}

A relevant question associated with the sesquinary catastrophe is: how is mass ejected from the minor planet in the first place? 

There are multiple processes which generate ejecta. The first is impacts, and the immediate vicinities ($\approx 1-4 r_{\rm Roche}$) of many white dwarfs are likely high-traffic bombardment zones. The reason is due to dust, pebbles and possibly boulders from distances of tens of au being pulled in by radiation \citep{veretal2022} and potentially magnetism \citep{hogetal2021,zhaetal2021,zhoetal2024}, as well as minor planets which have been thrust inwards due to existing massive perturbers (which can be as as high as stellar masses or as low as Luna masses; \citealt*{verros2023}) both inside or outside of the system.

This inwards scattering process has been extensively investigated with both single perturbers \citep{bonetal2011,debetal2012,frehan2014,picetal2017,antver2019,veretal2021,jinetal2023,mcdver2023,veretal2023,kuretal2024,rodlai2024} and multiple ones \citep{bonver2015,hampor2016,petmun2017,steetal2017,musetal2018,smaetal2018,smaetal2021,ocoetal2022,stoetal2022,trietal2022}. These studies document a wide variety of architectures and dynamics which illustrate how minor planets can be gravitationally perturbed into the close vicinity of white dwarfs. Further, the above reference lists do not even account for the perturbations on exo-Oort cloud comets around white dwarfs \citep[e.g.][]{ocoetal2023,pharei2024}.

Besides impacts, other ways of generating mass ejecta are through radiation-induced YORP-style break-up \citep{veretal2022}, sublimation \citep{broetal2017,mcdver2021}, thermal cracking \citep{sheser2023} and rotational surface shedding \citep{jacsch2011,zhaetal2017,zhamic2020,jacetal2022}. Partial mass shedding may occur through spin-up, or, alternatively, if the minor planet is composed of a mixture of high-density, low-density, high-strength and low-strength regions, each with their own Roche radius, only some of which are breached \citep{veretal2017,duvetal2020,broetal2023}.

\subsection{Refining the collisional timescale}

Another relevant question is: how accurate of a proxy is $\tau$ for the collisional timescale? We can guess the answer by considering the initial orbits of the ejecta. These orbits could vary significantly depending on how the ejecta is generated, with the ejecta methods described in the last paragraph.

Due to the frequency of conjunctions, ejecta on very similar initial orbits to the parent body will be the slowest to collide, whereas those on initially different orbits will collide more quickly. The reason is due to the timing of potential collisions: if the ejecta and the parent body initially have very similar orbital periods, then their synodic periods would be long and collisions would proportionally take a longer time. Alternatively, for initially more divergent orbits, synodic periods are shorter and hence collisions would occur in a shorter amount of time, on average. Note that this argument assumes that eccentricities and inclinations of ejecta are largely inherited from the parent body, rather than being enhanced in the ejection event. Another consequence of different orbital distances is the varying time necessary for differential precession, with more divergent orbits taking less time to precess out of alignment. This situation does not make collisions more frequent for debris on more dissimilar orbits, but it does significantly shorten the minimum time between the initial ejection of debris and first high-velocity re-impacts that can trigger runaway erosion.

Further, by way of comparison to the solar system, if the ejecta's initial orbit differs too much from that of its parent satellite, then the ejecta will become heliocentric and not ever re-impact the satellite. In contrast, the gravitational well of a white dwarf is often strong enough to force the ejecta to remain in the system; this principle is why small planetary perturbers, which cannot eject debris from the system, may be the most efficient drivers of white dwarf pollution \citep{verros2023}. Therefore, the allowable parameter space for ejected mass to return to parent bodies in white dwarf planetary systems is greater than in e.g. the Martian or Saturnian systems.

Understanding how similar the ejecta orbit is to its parent body necessitates detailed modelling of each process which creates the ejecta, which is beyond the scope of this study. Nevertheless, we can speculate based on exit velocity and orientation considerations. Thermal cracking and rotational breakup, even just as the trigger for surface mass shedding, is unlikely to significantly alter the initial orbits of ejecta. Alternatively, sublimation, as evidenced from solar system observations, may produce comet-like jets of dust and ions, and the resulting ejecta orbits may differ more significantly.

However, the greatest initial variation in ejecta orbits is likely to be produced by initial impacts from incoming objects which originate in the outer planetary system. One formulation linking the parent body orbital elements to the ejecta's post-impact initial orbital elements as a function of impact speed and orientation can be found in \cite{jacetal2014}. Their formalism suggests that the orbital elements of the ejecta's post-impact orbit could vary substantially from those of the parent body, allowing subsequent differential precession to quickly mix their orbits.

\subsection{Tidal inspiral}

A further relevant question is: once a rubble pile has been emplaced onto a $T\approx 5-25$~hr orbit which completely or partially avoids the rubble-pile Roche radius, then how long will that object survive before being tidally engulfed? 

The majority of dedicated tidal studies of white dwarf planetary systems have focussed on the tidal interactions of large planets rather than minor planets \citep{verful2019,verful2020,becetal2023,becetal2025}. The other tidal studies which featured smaller bodies \citep{ocolai2020,lietal2025a,lietal2025b} focussed on other (important) aspects, such as the origin and distribution of minor bodies in $T\approx 5-25$~hr orbits. \cite{veretal2019} claimed that with their tidal prescriptions, minor planets are too small to experience significant tidal inspiral when close to the Roche radius. 

Probably the more pertinent question is then: how does tidal circularisation and shrinkage affect the sesquinary catastrophe even {\it while} the pericentre remains beyond the Roche radius? Equations 11a and 18 of \cite{ocolai2020} and equation 13 of \cite{lietal2025a} all provide fiducial tidal timescales well in excess of $10^7$~yr for minor planets with $e>0.99$ and pericentres close to the Roche radius\footnote{Other means of circularisation, such as through drag from pre-existing dust or gas \citep{maletal2021}, can yield different timescales.}. Sesquinary catastrophe occurrence timescales of $\approx 10^2-10^5$~yr are much quicker. Further, the catastrophe is likely to take hold before tidal circularisation has been completed, and soon after a highly eccentric minor planet's orbit shrinks and circularises to a point where apsidal precession can differentiate ejecta orbits from that of the minor planet.

\section{Summary}\label{summary}

Complex photometric transit signatures of planetary debris orbiting white dwarfs, as well as theoretical studies of minor planets approaching these evolved stars from au-scale distances, indicate that the region corresponding to orbital periods $T\approx 5-25$~hr ($a \approx 1-4\,r_{\rm Roche}$) is a rich hub of dynamical activity. These orbital periods do not guarantee that a progenitor minor planet has reached the white dwarf's Roche radius, calling into question potential destruction mechanisms and complicating the replenishment rate and lifetime estimates of the resulting dusty and gaseous debris discs.

Here we have demonstrated that a potential destruction mechanism in this region is the sesquinary catastrophe \citep{cuketal2023}, a collision-based mechanism which effectively features a minor planet hitting itself with ejected fragments that return with destructively fast speeds on excited orbits. The minor planet orbit needs to be at least slightly eccentric (Fig. \ref{gvplots}) for this phenomenon to occur, a threshold which is easily reached if the orbit is gradually circularised from au-scale distances after the giant branch phases of evolution. The timescale for the sesquinary catastrophe to occur is expected to be longer than 10-100 yrs but much shorter than Myrs (Figs. \ref{gvplots} and \ref{tauplot1}), placing a useful upper bound on the destruction epoch for a minor planet whose pericentre never quite, or only partially, reaches the tidal fragmentation region.

\section*{Acknowledgements}

We thank the reviewer for perceptive and helpful comments that have improved the manuscript.

\section*{Data Availability}

All data presented in this paper is available upon reasonable request to the authors.

\label{lastpage}

\begin{thebibliography}{99}

 
\bibitem[Alvarellos et al.(2005)]{alvetal2005} Alvarellos, J.~L., Zahnle, K.~J., Dobrovolskis, A.~R., Hamill, P. \ 2005, Icarus, 
178, 1, 104. doi:10.1016/j.icarus.2005.04.017

\bibitem[Alvarellos et al.(2008)]{alvetal2008} Alvarellos, J.~L., Zahnle, K.~J., Dobrovolskis, A.~R., Hamill, P.\ 2008, Icarus, 
194, 2, 636. doi:10.1016/j.icarus.2007.09.025

\bibitem[Alvarellos et al.(2017)]{alvetal2017} Alvarellos, J.~L., Dobrovolskis, A.~R., Zahnle, K.~J., Hamill, P., Dones, L., Robbins, S.\ 2017, Icarus, 
284, 70. doi:10.1016/j.icarus.2016.10.028

\bibitem[Antoniadou \& Veras(2019)]{antver2019} Antoniadou, K.~I. \& Veras, D.\ 2019, A\&A, 
629, A126. doi:10.1051/0004-6361/201935996

\bibitem[Antoniadou \& Veras(2024)]{antver2024} Antoniadou, K.~I. \& Veras, D.\ 2024, A\&A, 
690, A249. doi:10.1051/0004-6361/202451714

\bibitem[Aungwerojwit et al.(2024)]{aunetal2024} Aungwerojwit, A. et al.\ 2024, MNRAS, 
530, 1, 117. doi:10.1093/mnras/stae750

\bibitem[Ballering et al.(2022)]{baletal2022} Ballering, N.~P., Levens, C.~I., Su, K.~Y.~L., Cleeves, L.~I.\ 2022, ApJ, 
939, 2, 108. doi:10.3847/1538-4357/ac9a4a

\bibitem[Batygin \& Adams(2013)]{batada2013} Batygin, K. \& Adams, F.~C.\ 2013, ApJ, 
778, 2, 169. doi:10.1088/0004-637X/778/2/169

\bibitem[Becker et al.(2023)]{becetal2023} Becker, J., Seligman, D.~Z., Adams, F.~C., Styczinski, M~.J.\ 2023, ApJL, 
945, 2, L24. doi:10.3847/2041-8213/acbe44

\bibitem[Becker et al.(2025)]{becetal2025} Becker, J., Vanderburg, A., \& Livesey, J.~R.\ 2025, ApJ, 
979, 2, 99. doi:10.3847/1538-4357/ada149

\bibitem[B{\'e}dard et al.(2020)]{bedetal2020} B{\'e}dard, A., Bergeron, P., Brassard, P., Fontaine, G.\ 2020, ApJ, 
901, 2, 93. doi:10.3847/1538-4357/abafbe

\bibitem[Bhattacharjee et al.(2025)]{bhaetal2025} Bhattacharjee, S. et al.\ 2025, 
Accepted for publication in PASP, arXiv:2502.05502. doi:10.48550/arXiv.2502.05502

\bibitem[Bierhaus et al.(2012)]{bieetal2012} Bierhaus, E.~B., Dones, L., Alvarellos, J.~L., Zahnle, K.\ 2012, Icarus, 
218, 1, 602. doi:10.1016/j.icarus.2011.12.011

\bibitem[Bonsor et al.(2011)]{bonetal2011} Bonsor, A., Mustill, A.~J., \& Wyatt, M.~C.\ 2011, MNRAS, 
414, 2, 930. doi:10.1111/j.1365-2966.2011.18524.x

\bibitem[Bonsor \& Veras(2015)]{bonver2015} Bonsor, A. \& Veras, D.\ 2015, MNRAS, 
454, 1, 53. doi:10.1093/mnras/stv1913

\bibitem[Briggs et al.(2018)]{brietal2018} Briggs, G.~P., Ferrario, L., Tout, C.~A., Wickramasinghe, D.~T.\ 2018, MNRAS, 
478, 1, 899. doi:10.1093/mnras/sty1150

\bibitem[Bromley \& Kenyon(2022)]{broken2022} Bromley, B.~C. \& Kenyon, S.~J.\ 2022, AJ, 
164, 6, 229. doi:10.3847/1538-3881/ac9301

\bibitem[Brouwers et al.(2022)]{broetal2022} Brouwers, M.~G., Bonsor, A., \& Malamud, U.\ 2022, MNRAS, 509, 2, 2404. doi:10.1093/mnras/stab3009

\bibitem[Brouwers et al.(2023)]{broetal2023} Brouwers, M.~G., Bonsor, A., \& Malamud, U.\ 2023, MNRAS, 
519, 2, 2646. doi:10.1093/mnras/stac3316

\bibitem[Brown et al.(2017)]{broetal2017} Brown, J.~C., Veras, D., \& G{\"a}nsicke, B.~T.\ 2017, MNRAS, 
468, 2, 1575. doi:10.1093/mnras/stx428

\bibitem[Coutu et al.(2019)]{couetal2019} Coutu, S., Dufour, P., Bergeron, P., Blouin, S., Loranger, E., Allard, N.~F., Dunlap, B.~H.  \ 2019, ApJ, 
885, 1, 74. doi:10.3847/1538-4357/ab46b9

\bibitem[{\'C}uk et al.(2023)]{cuketal2023} {\'C}uk, M., Hamilton, D.~P., Minton, D.~A., Stewart, S.~T. \ 2023, ApJ, 
957, 2, 62. doi:10.3847/1538-4357/acf613

\bibitem[Cunningham et al.(2025)]{cunetal2025} Cunningham, T. et al.\ 2025, MNRAS, 

\bibitem[Debes \& Sigurdsson(2002)]{debsig2002} Debes, J.~H. \& Sigurdsson, S.\ 2002, ApJ, 572, 1, 556. doi:10.1086/340291

\bibitem[Debes et al.(2012)]{debetal2012} Debes, J.~H., Walsh, K.~J., \& Stark, C.\ 2012, ApJ, 
747, 2, 148. doi:10.1088/0004-637X/747/2/148

\bibitem[Debes et al.(2025)]{debetal2025} Debes, J.~H. et al.\ 2025, , arXiv:2506.21224, accepted for publication in AJ. doi:10.48550/arXiv.2506.21224

\bibitem[Duvvuri et al.(2020)]{duvetal2020} Duvvuri, G.~M., Redfield, S., \& Veras, D.\ 2020, ApJ, 
893, 2, 166. doi:10.3847/1538-4357/ab7fa0

\bibitem[Farihi et al.(2009)]{faretal2009} Farihi, J., Jura, M., \& Zuckerman, B.\ 2009, ApJ, 
694, 2, 805. doi:10.1088/0004-637X/694/2/805

\bibitem[Farihi et al.(2022)]{faretal2022} Farihi, J. et al.\ 2022, MNRAS, 
511, 2, 1647. doi:10.1093/mnras/stab3475

\bibitem[Farihi et al.(2025)]{faretal2025} Farihi, J., Su, K.~Y.~L., Melis, C., et al.\ 2025, ApJL, 
981, 1, L5. doi:10.3847/2041-8213/adae88

\bibitem[Ferrario et al.(2015)]{feretal2015} Ferrario, L., de Martino, D., \& G{\"a}nsicke, B.~T.\ 2015, Solar System Reviews, 
191, 1-4, 111. doi:10.1007/s11214-015-0152-0

\bibitem[Fortin-Archambault et al.(2020)]{foretal2020} Fortin-Archambault, M., Dufour, P., \& Xu, S.\ 2020, ApJ, 
888, 1, 47. doi:10.3847/1538-4357/ab585a

\bibitem[French et al.(2025)]{freetal2025} French, R.~G. et al.\ 2025, Icarus, 
431, 116463. doi:10.1016/j.icarus.2025.116463

\bibitem[Frewen \& Hansen(2014)]{frehan2014} Frewen, S.~F.~N. \& Hansen, B.~M.~S.\ 2014, MNRAS, 
439, 3, 2442. doi:10.1093/mnras/stu097

\bibitem[G{\"a}nsicke et al.(2006)]{ganetal2006} G{\"a}nsicke, B.~T., Marsh, T.~R., Southworth, J., Rebassa-Mansergas, A. \ 2006, Science, 
314, 5807, 1908. doi:10.1126/science.1135033

\bibitem[G{\"a}nsicke et al.(2016)]{ganetal2016} G{\"a}nsicke, B.~T. et al.\ 2016, ApJL, 
818, 1, L7. doi:10.3847/2041-8205/818/1/L7

\bibitem[G{\"a}nsicke et al.(2019)]{ganetal2019} G{\"a}nsicke, B.~T., Schreiber, M.~R., Toloza, O., Gentile-Fusillo, N.~P., Koester, D., Manser, C.~J. \ 2019, Nature, 576, 61. doi:10.1038/s41586-019-1789-8

\bibitem[Gary et al.(2017)]{garetal2017} Gary, B.~L., Rappaport, S., Kaye, T.~G., Alonso, R., Hambschs, F.~J. \ 2017, MNRAS, 
465, 3, 3267. doi:10.1093/mnras/stw2921

\bibitem[Girven et al.(2012)]{giretal2012} Girven, J., Brinkworth, C.~S., Farihi, J., G{\"a}nsicke, B.~T., Hoard, D.~W., Marsh, T.~R., Koester, D. \ 2012, ApJ, 749, 154

\bibitem[Graham et al.(1990)]{graetal1990} Graham, J.~R., Matthews, K., Neugebauer, G., Soifer, B.~T. \ 1990, ApJ, 
357, 216. doi:10.1086/168907

\bibitem[Gurri et al.(2017)]{guretal2017} Gurri, P., Veras, D., \& G{\"a}nsicke, B.~T.\ 2017, MNRAS, 
464, 1, 321. doi:10.1093/mnras/stw2293

\bibitem[Hamers \& Portegies Zwart(2016)]{hampor2016} Hamers, A.~S. \& Portegies Zwart, S.~F.\ 2016, MNRAS, 
462, 1, L84. doi:10.1093/mnrasl/slw134

\bibitem[Hermes et al.(2025)]{heretal2025} Hermes, J.~J., Guidry, J.~A., Vanderbosch, Z.~P., Badenas-Agusti, M., Xu, S., Kao, M.~L., Rodriguez, A.~C., Hawkins, K.\ 2025, ApJ, 
980, 1, 56. doi:10.3847/1538-4357/ada5fd

\bibitem[Hogg et al.(2021)]{hogetal2021} Hogg, M.~A., Cutter, R., \& Wynn, G.~A.\ 2021, MNRAS, 
500, 3, 2986. doi:10.1093/mnras/staa3316

\bibitem[Iess et al.(2019)]{iesetal2019} Iess, L. et al.\ 2019, Science, 
364, 6445, aat2965. doi:10.1126/science.aat2965

\bibitem[Izquierdo et al.(2018)]{izqetal2018} Izquierdo, P. et al.\ 2018, MNRAS, 
481, 1, 703. doi:10.1093/mnras/sty2315

\bibitem[Jacobson \& Scheeres(2011)]{jacsch2011} Jacobson, S.~A. \& Scheeres, D.~J.\ 2011, Icarus, 
214, 1, 161. doi:10.1016/j.icarus.2011.04.009

\bibitem[Jackson et al.(2014)]{jacetal2014} Jackson, A.~P., Wyatt, M.~C., Bonsor, A., Veras, D. \ 2014, MNRAS, 
440, 4, 3757. 

\bibitem[Jackson et al.(2022)]{jacetal2022} Jackson, P.~M., Nakano, R., Kim, Y., Hirabayashi, M.\ 2022, PSJ, 
3, 1, 16. doi:10.3847/PSJ/ac4031

\bibitem[Jin et al.(2023)]{jinetal2023} Jin, Z., Li, D., \& Zhu, Z.-H.\ 2023, A\&A, 
674, A52. doi:10.1051/0004-6361/202345954

\bibitem[Jura(2003)]{jura2003} Jura, M.\ 2003, ApjL, 
584, 2, L91. doi:10.1086/374036

\bibitem[Kenyon \& Bromley(2017a)]{kenbro2017a} Kenyon, S.~J. \& Bromley, B.~C.\ 2017a, ApJ, 
844, 2, 116. doi:10.3847/1538-4357/aa7b85

\bibitem[Kenyon \& Bromley(2017b)]{kenbro2017b} Kenyon, S.~J. \& Bromley, B.~C.\ 2017b, ApJ, 
850, 1, 50. doi:10.3847/1538-4357/aa9570

\bibitem[Kjurkchieva et al.(2017)]{kjuetal2017} Kjurkchieva, D.~P., Dimitrov, D.~P., \& Petrov, N.~I.\ 2017, PASA, 
34, e032. doi:10.1017/pasa.2017.28

\bibitem[Koester et al.(2014)]{koeetal2014} Koester, D., G{\"a}nsicke, B.~T., \& Farihi, J.\ 2014, A\&A, 
566, A34. doi:10.1051/0004-6361/201423691

\bibitem[Kurban et al.(2024)]{kuretal2024} Kurban, A., Zhou, X., Wang, N., Huang, Y.-F., Wang, Y.-B., Nurmamat, N.\ 2024, ApJ, 
974, 1, 100. doi:10.3847/1538-4357/ad73d3

\bibitem[Li et al.(2021)]{lietal2021} Li, D., Mustill, A.~J., \& Davies, M.~B.\ 2021, MNRAS, 
508, 4, 5671. doi:10.1093/mnras/stab2949

\bibitem[Li et al.(2025a)]{lietal2025a} Li, Y., Bonsor, A., Shorttle, O., Rogers, L.~K.\ 2025a, MNRAS, 
537, 2, 2214. doi:10.1093/mnras/staf182

\bibitem[Li et al.(2025b)]{lietal2025b} Li, Y., Bonsor, A., Shorttle, O.\ 2025b, MNRAS In Press, arXiv:2506.20316

\bibitem[Makarov \& Veras(2019)]{makver2019} Makarov, V.~V. \& Veras, D.\ 2019, ApJ, 
886, 2, 127. doi:10.3847/1538-4357/ab4c95

\bibitem[Malamud et al.(2021)]{maletal2021} Malamud, U., Grishin, E., \& Brouwers, M.\ 2021, MNRAS, 
501, 3, 3806. doi:10.1093/mnras/staa3940

\bibitem[Malamud(2024)]{malamud2024} Malamud, U.\ 2024, Encyclopedia of Astrophysics, White dwarf systems: exoplanets and debris disks, arXiv:2403.07427. doi:10.48550/arXiv.2403.07427

\bibitem[Manser et al.(2019)]{manetal2019} Manser, C.~J. et al.\ 2019, Science, 
364, 6435, 66. doi:10.1126/science.aat5330

\bibitem[Manser et al.(2020)]{manetal2020} Manser, C.~J., G{\"a}nsicke, B.~T., Gentile Fusillo, N.~P., Ashley, R., Breedt, E., Hollands, M., Izquierdo, P, Pelisoli, I.\ 2020, MNRAS, 
493, 2, 2127. doi:10.1093/mnras/staa359

\bibitem[McCleery et al.(2020)]{mccetal2020} McCleery, J. et al.\ 2020, MNRAS, 
499, 2, 1890. doi:10.1093/mnras/staa2030

\bibitem[McDonald \& Veras(2021)]{mcdver2021} McDonald, C.~H. \& Veras, D.\ 2021, MNRAS, 
506, 3, 4031. doi:10.1093/mnras/stab1906

\bibitem[McDonald \& Veras(2023)]{mcdver2023} McDonald, C.~H. \& Veras, D.\ 2023, MNRAS, 
520, 3, 4009. doi:10.1093/mnras/stad382

\bibitem[Mustill et al.(2018)]{musetal2018} Mustill, A.~J., Villaver, E., Veras, D., Bonsor, A.\ 2018, MNRAS, 
476, 3, 3939. doi:10.1093/mnras/sty446

\bibitem[Nauenberg(1972)]{nauenberg1972} Nauenberg, M.\ 1972, ApJ, 
175, 417. doi:10.1086/151568

\bibitem[Nayak \& Asphaug(2016)]{nayasp2016} Nayak, M. \& Asphaug, E.\ 2016, Nature Communications, 
7, 12591. doi:10.1038/ncomms12591

\bibitem[Nayak et al.(2016)]{nayetal2016} Nayak, M., Nimmo, F., \& Udrea, B.\ 2016, Icarus, 
267, 220. doi:10.1016/j.icarus.2015.12.026

\bibitem[O'Brien et al.(2023)]{obretal2023} O'Brien, M.~W. et al.\ 2023, MNRAS, 
518, 2, 3055. doi:10.1093/mnras/stac3303

\bibitem[O'Connor \& Lai(2020)]{ocolai2020} O'Connor, C.~E. \& Lai, D.\ 2020, MNRAS, 
498, 3, 4005. doi:10.1093/mnras/staa2645

\bibitem[O'Connor et al.(2022)]{ocoetal2022} O'Connor, C.~E., Teyssandier, J., \& Lai, D.\ 2022, MNRAS, 
513, 3, 4178. doi:10.1093/mnras/stac1189

\bibitem[O'Connor et al.(2023)]{ocoetal2023} O'Connor, C.~E., Lai, D., \& Seligman, D.~Z.\ 2023, MNRAS, 
524, 4, 6181. doi:10.1093/mnras/stad2281

\bibitem[Oliveira da Rosa et al.(2024)]{olietal2024} Oliveira da Rosa, G., Kepler, S.~O., Soethe, L.~T.~T., Romero, A.~D., Bell, K.~J. \ 2024, ApJ, 
974, 2, 314. doi:10.3847/1538-4357/ad6987

\bibitem[Petrovich \& Mu{\~n}oz(2017)]{petmun2017} Petrovich, C. \& Mu{\~n}oz, D.~J.\ 2017, ApJ, 
834, 2, 116. doi:10.3847/1538-4357/834/2/116

\bibitem[Pham \& Rein(2024)]{pharei2024} Pham, D. \& Rein, H.\ 2024, MNRAS, 
530, 3, 2526. doi:10.1093/mnras/stae986

\bibitem[Pichierri et al.(2017)]{picetal2017} Pichierri, G., Morbidelli, A., \& Lai, D.\ 2017, A\&A, 
605, A23. doi:10.1051/0004-6361/201730936

\bibitem[Rappaport et al.(2016)]{rapetal2016} Rappaport, S., Gary, B.~L., Kaye, T., Vanderburg, A., Croll, B., Benni, P., Foote, J.\ 2016, MNRAS, 
458, 4, 3904. doi:10.1093/mnras/stw612

\bibitem[Rappaport et al.(2018)]{rapetal2018} Rappaport, S., Gary, B.~L., Vanderburg, A., Xu, S. Pooley, D., Mukai, K.\ 2018, MNRAS, 
474, 1, 933. doi:10.1093/mnras/stx2663

\bibitem[Rodet \& Lai(2024)]{rodlai2024} Rodet, L. \& Lai, D.\ 2024, MNRAS, 
527, 4, 11664. doi:10.1093/mnras/stad3905

\bibitem[Rogers et al.(2025)]{rogetal2025} Rogers, L.~K. et al.\ 2025, MNRAS, 
537, 1, L72. doi:10.1093/mnrasl/slae117

\bibitem[Shestakova \& Serebryanskiy(2023)]{sheser2023} Shestakova, L.~I. \& Serebryanskiy, A.~V.\ 2023, MNRAS, 
524, 3, 4506. doi:10.1093/mnras/stad2006

\bibitem[Smallwood et al.(2018)]{smaetal2018} Smallwood, J.~L., Martin, R.~G., Livio, M., Lubow, S.~H.\ 2018, MNRAS, 
480, 1, 57. doi:10.1093/mnras/sty1819

\bibitem[Smallwood et al.(2021)]{smaetal2021} Smallwood, J.~L., Martin, R.~G., Livio, M., Veras, D.\ 2021, MNRAS, 
504, 3, 3375. doi:10.1093/mnras/stab1077

\bibitem[Spalding et al.(2018)]{spaetal2018} Spalding, C., Marx, N.~W., \& Batygin, K.\ 2018, AJ, 
155, 4, 167. doi:10.3847/1538-3881/aab43a

\bibitem[Stephan et al.(2017)]{steetal2017} Stephan, A.~P., Naoz, S., \& Zuckerman, B.\ 2017, ApJL, 
844, 2, L16. doi:10.3847/2041-8213/aa7cf3

\bibitem[Sterne(1939)]{sterne1939} Sterne, T.~E.\ 1939, MNRAS, 
99, 451. doi:10.1093/mnras/99.5.451

\bibitem[Stewart \& Leinhardt(2012)]{stelei2012} Stewart, S.~T. \& Leinhardt, Z.~M.\ 2012, ApJ, 
751, 1, 32. doi:10.1088/0004-637X/751/1/32

\bibitem[Stock et al.(2022)]{stoetal2022} Stock, K., Veras, D., Cai, M.~X., Spurzem, R., Portegies Zwart, S.\ 2022, MNRAS, 
512, 2, 2460. doi:10.1093/mnras/stac602

\bibitem[Swan et al.(2024)]{swaetal2024} Swan, A., Farihi, J., Su, K.~Y.~L., et al.\ 2024, MNRAS, 529, 1, L41. doi:10.1093/mnrasl/slad198

\bibitem[Trevascus et al.(2021)]{treetal2021} Trevascus, D., Price, D.~J., Nealon, R., Liptai, D., Manser, C.~J.\ 2021, MNRAS, 
505, 1, L21. doi:10.1093/mnrasl/slab043

\bibitem[Trierweiler et al.(2022)]{trietal2022} Trierweiler, I.~L., Doyle, A.~E., Melis, C., Walsh, K.~J., Young, E.~D.\ 2022, ApJ, 
936, 1, 30. doi:10.3847/1538-4357/ac86d5

\bibitem[van Lieshout et al.(2018)]{vanetal2018} van Lieshout, R., Kral, Q., Charnoz, S., Wyatt, M.~C., Shannon, A.\ 2018, MNRAS, 
480, 2, 2784. doi:10.1093/mnras/sty1271

\bibitem[Vanderburg et al.(2020)]{WD1856} Vanderburg, A., Rappaport, S.~A., Xu, S., et al.\ 2020, Nature, 
585, 7825, 363. doi:10.1038/s41586-020-2713-y

\bibitem[Vanderbosch et al.(2020)]{vanbosetal2020} Vanderbosch, Z. et al.\ 2020, ApJ, 
897, 2, 171. doi:10.3847/1538-4357/ab9649

\bibitem[Vanderbosch et al.(2021)]{vanbosetal2021} Vanderbosch, Z.~P. et al.\ 2021, ApJ, 
917, 1, 41. doi:10.3847/1538-4357/ac0822

\bibitem[Vanderburg et al.(2015)]{vanetal2015} Vanderburg, A. et al.\ 2015, 
Nature, 
526, 7574, 546. doi:10.1038/nature15527

\bibitem[Vanderburg \& Rappaport(2018)]{vanrap2018} Vanderburg, A. \& Rappaport, S.~A.\ 2018, 
37. doi:10.1007/978-3-319-55333-7\_37

\bibitem[Veras et al.(2016)]{veretal2016} Veras, D., Marsh, T.~R., \& G{\"a}nsicke, B.~T.\ 2016, MNRAS, 
461, 2, 1413. doi:10.1093/mnras/stw1324

\bibitem[Veras et al.(2017)]{veretal2017} Veras, D., Carter, P.~J., Leinhardt, Z.~M., et al.\ 2017, MNRAS, 
465, 1, 1008. doi:10.1093/mnras/stw2748

\bibitem[Veras \& Fuller(2019)]{verful2019} Veras, D. \& Fuller, J.\ 2019, MNRAS, 
489, 2, 2941. doi:10.1093/mnras/stz2339

\bibitem[Veras et al.(2019)]{veretal2019} Veras, D., Efroimsky, M., Makarov, V.~V., et al.\ 2019, MNRAS, 
486, 3, 3831. doi:10.1093/mnras/stz965

\bibitem[Veras \& Fuller(2020)]{verful2020} Veras, D. \& Fuller, J.\ 2020, MNRAS, 
492, 4, 6059. doi:10.1093/mnras/staa309

\bibitem[Veras \& Heng(2020)]{verhen2020} Veras, D. \& Heng, K.\ 2020, MNRAS, 
496, 2, 2292. doi:10.1093/mnras/staa1632

\bibitem[Veras \& Kurosawa(2020)]{verkur2020} Veras, D. \& Kurosawa, K.\ 2020, MNRAS, 
494, 1, 442. doi:10.1093/mnras/staa621

\bibitem[Veras et al.(2020)]{veretal2020} Veras, D., McDonald, C.~H., \& Makarov, V.~V.\ 2020, MNRAS, 
492, 4, 5291. doi:10.1093/mnras/staa243

\bibitem[Veras et al.(2021)]{veretal2021} Veras, D., Georgakarakos, N., Mustill, A.~J., Malamud, U., Cunningham, T., Dobbs-Dixon, I.\ 2021, MNRAS, 
506, 1, 1148. doi:10.1093/mnras/stab1667

\bibitem[Veras(2021)]{veras2021} Veras, D.\ 2021, Oxford Research Encyclopedia of Planetary Science, Planetary Systems Around White Dwarfs, 1. doi:10.1093/acrefore/9780190647926.013.238

\bibitem[Veras et al.(2022)]{veretal2022} Veras, D., Birader, Y., \& Zaman, U.\ 2022, MNRAS, 
510, 3, 3379. doi:10.1093/mnras/stab3490

\bibitem[Veras \& Rosengren(2023)]{verros2023} Veras, D. \& Rosengren, A.~J.\ 2023, MNRAS, 
519, 4, 6257. doi:10.1093/mnras/stad130

\bibitem[Veras et al.(2023)]{veretal2023} Veras, D., Georgakarakos, N., \& Dobbs-Dixon, I.\ 2023, MNRAS, 
518, 3, 4537. doi:10.1093/mnras/stac3274

\bibitem[Veras et al.(2024)]{veretal2024} Veras, D., Mustill, A.~J., \& Bonsor, A.\ 2024, Reviews in Mineralogy and Geochemistry, Volume 90: The Evolution and Delivery of Rocky Extra-Solar Materials to White Dwarfs, Edited by Natalie Hinkel, Keith Putirka, and Siyi Xu; 90, 1, 141. doi:10.2138/rmg.2024.90.05

\bibitem[Ward et al.(1976)]{waretal1976} Ward, W.~R., Colombo, G., \& Franklin, F.~A.\ 1976, Icarus, 
28, 4, 441. doi:10.1016/0019-1035(76)90117-2

\bibitem[Williams et al.(2024)]{wiletal2024} Williams, J.~T., G{\"a}nsicke, B.~T., Swan, A., O'Brien, M.~W., Izquierdo, P., Cutolo, A.~M., Cunningham, T. \ 2024, A\&A, 
691, A352. doi:10.1051/0004-6361/202450509

\bibitem[Wyatt et al.(2014)]{wyaetal2014} Wyatt, M.~C., Farihi, J., Pringle, J.~E., Bonsor, A.\ 2014, MNRAS, 
439, 4, 3371. doi:10.1093/mnras/stu183

\bibitem[Xu et al.(2020)]{xuetal2020} Xu, S., Lai, S., \& Dennihy, E.\ 2020, ApJ, 
902, 2, 127. doi:10.3847/1538-4357/abb3fc

\bibitem[Zahnle et al.(2008)]{zahetal2008} Zahnle, K., Alvarellos, J.~L., Dobrovolskis, A., Hamill, P.\ 2008, Icarus, 
194, 2, 660. doi:10.1016/j.icarus.2007.10.024

\bibitem[Zhang et al.(2017)]{zhaetal2017} Zhang, Y. et al.\ 2017, Icarus, 
294, 98. doi:10.1016/j.icarus.2017.04.027

\bibitem[Zhang \& Michel(2020)]{zhamic2020} Zhang, Y. \& Michel, P.\ 2020, A\&A, 
640, A102. doi:10.1051/0004-6361/202037856

\bibitem[Zhang et al.(2021)]{zhaetal2021} Zhang, Y., Liu, S.-F., \& Lin, D.~N.~C.\ 2021, ApJ, 
915, 2, 91. doi:10.3847/1538-4357/ac00ae

\bibitem[Zhou et al.(2024)]{zhoetal2024} Zhou, W.-H., Liu, S.-F., \& Lin, D.~N.~C.\ 2024, A\&A, 
687, A107. doi:10.1051/0004-6361/202449271

\bibitem[Zuckerman \& Becklin(1987)]{zucbec1987} Zuckerman, B. \& Becklin, E.~E.\ 1987, Nature, 
330, 6144, 138. doi:10.1038/330138a0

\bibitem[Zuckerman et al.(2010)]{zucetal2010} Zuckerman, B., Melis, C., Klein, B., Koester, D., Jura, M.\ 2010, ApJ, 
722, 1, 725. doi:10.1088/0004-637X/722/1/725


\end{thebibliography}
\end{document}